# Orbital-Resolved Partial Charge Transfer from the Methoxy Groups of Substituted Pyrenes in Complexes with Tetracyanoquinodimethane - a NEXAFS Study

K. Medjanik, S. A. Nepijko, H. J. Elmers, G. Schönhense*
*Institut für Physik, Johannes Gutenberg-Universität Mainz, Germany*

P. Nagel, M. Merz, S. Schuppler
*Karlsruhe Institute of Technology (KIT), Institut für Festkörperphysik, 76021 Karlsruhe, Germany*

D. Chercka, M. Baumgarten, and K. Müllen
*Max-Planck-Institut für Polymerforschung, Mainz, Germany*

**Abstract**

It is demonstrated that the near-edge X-ray absorption fine structure (NEXAFS) provides a powerful local probe of functional groups in novel charge transfer (CT) compounds. Microcrystals of tetra- and hexamethoxypyrene as donors with the strong acceptor tetracyanoquinodimethane ($TMP_x/HMP_x$ - $TCNQ_y$) were grown from solution via vapour diffusion in different stoichiometries x:y = 1:1, 1:2 and 2:1. Owing to the element specificity of NEXAFS, the oxygen and nitrogen K-edge spectra are direct spectroscopic fingerprints of the donating and accepting moieties. The orbital selectivity of the NEXAFS resonances allows to precisely elucidate the participation of specific orbitals in the charge-transfer process. In the present case charge is transferred from methoxy-orbitals $2e$ ($\pi^*$) and $6a_1$ ($\sigma^*$) to the cyano-orbitals $b_{3g}$ and $a_u$ ($\pi^*$) and - to a weaker extent - to $b_{1g}$ and $b_{2u}$ ($\sigma^*$). The occupation of $2e$ reflects the anionic character of the methoxy groups. Surprisingly, the charge transfer increases with increasing HMP content of the complex. As additional indirect signature, all spectral features of the donor and acceptor are shifted to higher and lower photon energies, respectively. Providing quantitative access to the relative occupation of specific orbitals, the approach constitutes the most direct probe of the charge-transfer mechanism in organic salts found so far. Although demonstrated for the specific example of pyrene-derived donors with the classical acceptor TCNQ, the method is very versatile and can serve as routine probe for novel CT-complexes on the basis of functionalized polycyclic aromatic hydrocarbons.

# 1. Introduction

## 1.1 Functionalized Polycyclic Aromatic Hydrocarbons With Strong Acceptor or Donor Character

In the field of molecular electronics conjugated organic molecules have received intense attention as n-type or p-type semiconductors. In particular, charge transfer (CT) compounds of molecules with tailored donor and acceptor character provide a vast multitude of design possibilities. Understanding the electronic structure of this class of materials as well as their metal-organic interfaces is crucial for designing specific electrical properties. In particular, large planar polycyclic aromatic hydrocarbon molecules with different functional groups at the periphery have recently been studied intensively. It was shown that their electronic properties can be tailored for strong electron acceptor or donor character [1]. Novel chemical synthesis routes as described for the coronene case by Rieger et al. [2] paved the way towards the design of a new class of donor and acceptor molecules both based on the same parent molecule. Indeed, the UHV co-deposited donor hexamethoxycoronene and acceptor coronene-hexaone form a weak CT complex as recently shown using photoemission and infrared spectroscopy [3].

Likewise, the pyrene molecule can be functionalized at its periphery by adding either methoxy- or keto-groups thus yielding moderate donors or strong acceptors, respectively [4]. The donor moieties are of particular interest because their sizes are similar to that of the classical strong acceptor *7,7,8,8-tetracyano-p-quinodimethane*, (TCNQ, $C_{12}N_4H_4$), see Fig.1a. In *4,5,9,10-tetramethoxypyrene* (TMP, $C_{20}H_{18}O_4$) the four functional groups lower the ionization potential (IP) to 5.47 eV [4], indicating a strongly increased donor character in comparison with the parent molecule pyrene (IP = 7.41 eV [5]). In *2,4,5,7,9,10-hexamethoxypyrene* (HMP, $C_{22}H_{22}O_6$) IP is further lowered to 5.17 eV, as measured by cyclovoltammetry. It is thus interesting to investigate the possible formation of new CT complexes based on these donor moieties, in particular in compounds with TCNQ. For the present study, new crystallographic phases with different stoichiometries of TCNQ and HMP as well as TMP have been grown from solution.

Thin films of the TMP-TCNQ complex on atomically clean gold surfaces have been produced by UHV co-deposition and investigated by ultraviolet photoelectron spectroscopy (UPS) and scanning tunnelling spectroscopy (STS) [6]. New reflexes in X-ray diffraction from these films gave evidence of a crystallographic phase different from pure TCNQ and TMP. The reflex positions in θ-2θ scans of the thin film samples are compatible with the results of the 3D-analysis of the solution-grown crystallites. Infrared spectra revealed a red-shift of the CN stretching vibration frequency by 7 cm$^{-1}$, indicating a charge transfer of about 0.3e. Shifts in the level positions of the frontier orbitals as visible in UPS and STS could be qualitatively interpreted on the basis of density functional theory calculations. The mixed-stack phases of the pyrene-derived donors TMP and HMP with TCNQ thus constitute a new class of charge transfer complexes.

Despite of the first results on the crystallographic structure of the solution-grown crystallites [4] and the mainly spectroscopic results on the UHV-deposited thin films [6], several important questions still remain unsolved. In particular, there is lack of information on the electronic structure of the solution-grown crystals because the unavoidable surface contaminations and the small size of the crystallites were prohibitive for UPS and STS analyses. The probing depth is only 2-3 molecular layers in UPS and only one molecular layer in STS. In the present work, we have employed X-ray absorption spectroscopy. It provides a larger probing depth in the order of 5 nm. Thus, surface contaminations are expected to be much less severe than in UPS or STS.

**1.2 Charge Transfer Salts Studied via Near-Edge X-ray Absorption Fine Structure (NEXAFS) Spectroscopy**

In NEXAFS measurements an electron is excited from a core level to an empty or partially unoccupied valence electronic state. As this technique gives direct access to the unoccupied density of states, it should be sensitive to the formation of new hole states due to charge transfer in a donor-acceptor complex [7]. By selection of a specific atom via its X-ray absorption edge, the electronic structure in the vicinity of this atom (e.g. in a functional group attached to a large molecule) is probed. For the present experiment the nitrogen 1s and oxygen 1s core levels should be ideal, because these atomic species are located exclusively on the acceptor site (N in the cyano-groups of TCNQ) or the donor site (O in the methoxy-groups of TMP and HMP). This highly specific excitation should provide information on the local electronic structure and thus on local chemical functionalities. We cannot expect carbon K-edge NEXAFS to be particularly useful due to the many non-equivalent C-atoms in both the donor and acceptor ring systems. Nitrogen and oxygen K-edge NEXAFS require monochromatic, tuneable synchrotron radiation in the soft X-ray range. In the present study we exploited soft X-rays from the WERA beamline at ANKA, Karlsruhe. In the total electron yield mode employed, the drain current from the sample is detected. The electron emission yield originates from a subsequent Auger process that neutralizes the core hole and leads to the emission of Auger electrons and slow secondary electrons. The information depth is about 5 nm in total yield mode.

The NEXAFS method is described in detail in [8]. Fraxedas et al. [9] performed a NEXAFS study of the classical charge transfer salt *tetrathiafulvalene* (TTF, $C_6S_4H_4$) – TCNQ. This compound forms parallel segregated stacks of donors (TTF) and acceptors (TCNQ). Results have been compared with a first-principles calculation of the unoccupied and partially occupied electronic states of the pure materials and the charge transfer compound. Later, Sing et al. [10] studied the same system with particular focus on the renormalized band widths observed in UPS [11,12] for the same compound. By variation of the angle of photon incidence, the symmetry of the observed orbitals for TCNQ was probed and information on molecular orientation was gained [10]. None of these papers addressed the issue of a possible change of the unoccupied density of states upon formation of the CT complex.

In the present paper we present NEXAFS results for solution-grown 3D crystallites of the complexes HMP-TCNQ and TMP-TCNQ in different stoichiometric mixtures. For comparison, spectra of the pure donors and acceptors have been taken under the same conditions. The study revealed two different signatures of a charge transfer in these new compounds. Strong changes in the intensity of the oxygen pre-edge features for different stoichiometries are a fingerprint of the occurrence of additional hole states at the donor sites. This pre-edge feature is absent in the spectrum of pure donor material, giving evidence of full anionicity of the methoxy group in HMP and TMP. Complementary, two prominent resonances of TCNQ are quenched, indicating partial filling of these states. In addition, characteristic energy shifts of the donor and acceptor NEXAFS resonances in opposite directions are a further consequence of the charge transfer.

## 2. Crystal Growth of CT Compounds from Solution via Vapour Diffusion

In 1962 Melby et al. [13] have grown pyrene-TCNQ in solution and found a weak CT complex; later Amano et al. [14] have grown diaminopyrene-TCNQ in acetonitrile $CH_3CN$ and found an ionic CT salt. In first experiments we just looked at the colour change of concentrated solution of the donor-acceptor moieties providing a broad CT absorption band in the visible range with maximum at about 600 nm.

Crystals were grown by vapour diffusion of hexane into a dichloromethane solution (5 ml, $6,2*10^{-3}$ mol/l) of the components. Solutions with donor-acceptor mixtures of 1:2, 1:1 and 2:1 stoichiometry were prepared. The components were combined in a glas vial (V = 7 cc, 1,5 cm diameter) and dissolved under sonication. Vapour diffusion assisted crystallizations were performed in a gas-tight chamber (V = 120 cc), filled with 15 ml hexane. The vial containing the solution was placed inside the chamber which was sealed for 4 days.

The crystallites have sizes in the range from several 10μm to several 100μm. X-ray diffraction analysis of TMP-TCNQ microcrystal fractions revealed a mixed-stack geometry as shown in Fig. 1b. [4] The complex appears black, whereas crystallites of the pure donors HMP or TMP are colourless transparent and TCNQ is transparent with light green colour. Optical microscopy revealed that there is a mixture of dark and transparent crystals in the vial, examples, see Fig. 1c and d. The admixture of bright crystals is maximum in the $HMP_2$-$TCNQ_1$-stoichiometry. For the 1:1 stoichiometry the admixture is smaller and for the 1:2 stoichiometry there are no transparent crystals. The different phases could easily be distinguished by their colour and the crystal fractions could thus be separated using a micromanipulator under the optical microscope. The NEXAFS spectra revealed that the transparent crystals show no nitrogen K-edge signal. This proves that they contain no TCNQ and consist of pure donor material. Obviously, HMP or TMP microcrystals form during the vapour diffusion process in coexistence with the complex.

## 3. Nitrogen and Oxygen K-edge NEXAFS Spectra

NEXAFS spectra of the different complexes and of pure donor and acceptor molecules have been taken at the WERA beamline of ANKA. The dark fractions of the solution-grown microcrystals were deposited on carbon tape, being a suitable holder for such samples. The results for the nitrogen and oxygen K-edge spectra are summarized in Fig. 2, 3, and 5. The spectral features (A-E for nitrogen and F-K for oxygen) were quantitatively analyzed by a multi-peak fit routine. The partial spectra resulting from the fit are shown as thin lines, circles denote the measured spectra.

As nitrogen is present only in the cyano-groups of TCNQ, its edge fine structure is a fingerprint of the acceptor. Likewise, oxygen is only contained in the methoxy-groups and thus its spectrum represents a local probe in the functional group of the donor. The abscissa is identical for all spectra shown. Clearly, the absolute total yield of the N K-edge spectra grows with increasing TCNQ content of the compounds, cf. sequence of spectra a, b, c and d, in Fig.2, the latter taken for pure TCNQ. Vice versa, the intensity of the oxygen features (Fig. 3) decreases in the sequence of spectra e (taken for pure HMP), f, g and h. This proves that the relative donor-acceptor concentrations in the solution show up in the stoichiometry of the 3D crystallites as well.

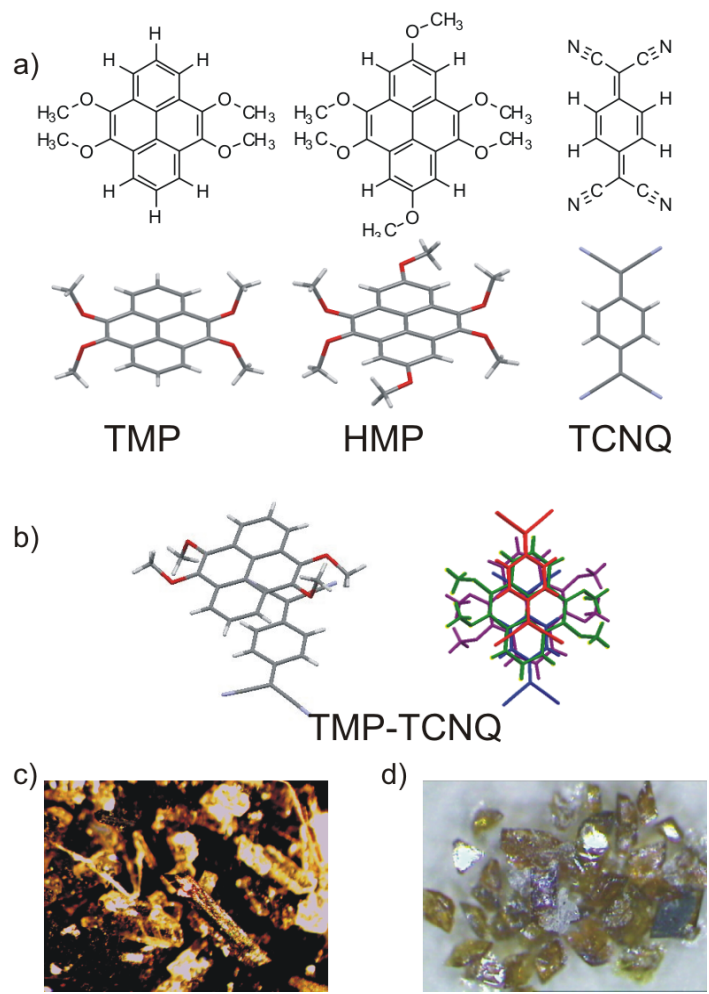

**Figure 1:** Molecular structures (a) and structure of the $TMP_1$-$TCNQ_1$ complex as obtained from X-ray diffraction (b); the second plot shows a colour-coded top view of the molecular arrangement in four adjacent layers. Optical microscopy of $HMP_2$-$TCNQ_1$ (c) and $TMP_1$-$TCNQ_1$ (d) shows coexistence of dark and bright microcrystals (field of view 1mm horiz.).

## 3.1 Nitrogen K-edge NEXAFS

The spectrum for the nitrogen K-edge was deconvoluted into signals A-E using the multi-peak fit routine, assuming a mixed Gaussian/Lorentzian function accounting for a finite energy resolution of 0.2 eV and a life-time broadening of 0.4 eV. Fit results are plotted in Fig.2 as thin curves below the spectra. The sum of all partial spectra is shown as full curve through the data points (circles). In all cases the fit curve perfectly reproduces the data points, giving evidence of a high reliability of the partial spectra. NEXAFS spectra of pure TCNQ have been analyzed in Ref. [9]. Here we briefly recall the peak assignment. The first signal (A) is a $\pi^*$–type resonance (resonant transition 1s→$\pi^*$) and originates from the lower $a_u$ and $b_{1u}$ orbitals. For the free molecule these lie at 2.55 eV and 2.65 eV above the LUMO minimum. At photon energies in resonance with such an allowed dipole transition, a large increase in excitation cross section is observed. The orbitals $a_u$ and $b_{1u}$ originate from the degenerate pair of lowest empty $\pi^*$-orbitals ($e_{2u}$) of the benzene core, which only slightly delocalize towards the cyano-group in TCNQ. We denote transition A as *N1s→$a_u$,$b_{1u}$[$\pi^*$(ring)]*. Signals B and C originate from p-type unoccupied orbitals located in the cyano-group and are therefore higher in intensity due to the larger overlap with the N1s wavefunction in the dipole matrix element. B is a $\sigma^*$-type resonance originating from the $b_{1g}$ and $b_{2u}$ orbitals that belong to the four symmetry adapted combinations of in-plane orbitals of the CN groups. The corresponding transition B is denoted as ***N1s→$b_{1g},b_{2u}$ [σ*(C≡N)]***. Signal C is of π*-type and derives from the $b_{3g}$ and $a_u$ orbitals, i.e. ***N1s→$b_{3g},a_u$[π*(C≡N)]***. Finally, signal D corresponds to the highest π*-type orbital of benzene ($b_{2g}$) that is delocalized over the whole TCNQ molecule, ***N1s→$b_{2g}$[π*(ring)]***. In summary, the unoccupied states involved in the weaker transitions A and D are essentially delocalized on the benzene ring, whereas states involved in transitions B and C largely consist of the σ*- and π*-orbitals of the cyano-group. The weak feature E and higher lying weak signals (not shown) originate from delocalized σ-type orbitals containing σ*(C-C), σ*(C-H) and σ*(C≡N) contributions.

Unlike the TCNQ-TTF case [9], where signals B and C merge to a single intense peak, they can be well separated by the fit (Fig. 2 a – c). In spectrum (d), showing the result for pure TCNQ powder, signal B appears as a low-energy shoulder. We have checked the order of signals B and C by taking spectra for thin-film samples of TCNQ (not shown here): At grazing incidence (75°) the π*-type signal (C) gains intensity, whereas at normal incidence (0°) signal (B) with σ*-symmetry has its intensity maximum. This is in agreement with the angular-dependent measurements of Sing et al. [10].

With increasing HMP content (sequence d, c, b, a) the signal intensities B and C drop substantially. For $HMP_2$-$TCNQ_1$ (spectrum a) signals B and C appear as a double peak with a separation of about 190 meV. This intensity drop is an indication of charge being transferred into these unoccupied orbitals of the acceptor molecule. Moreover, significant shifts of the resonance positions towards lower hν occur as function of the HMP content (see vertical lines). We will return to these points in the context of a complementary behaviour found for the oxygen pre-edge signal of the donor molecule. For the TTF-TCNQ complex, only small variations in N K-edge NEXAFS spectra in comparison with pure, neutral TCNQ have been observed [9].

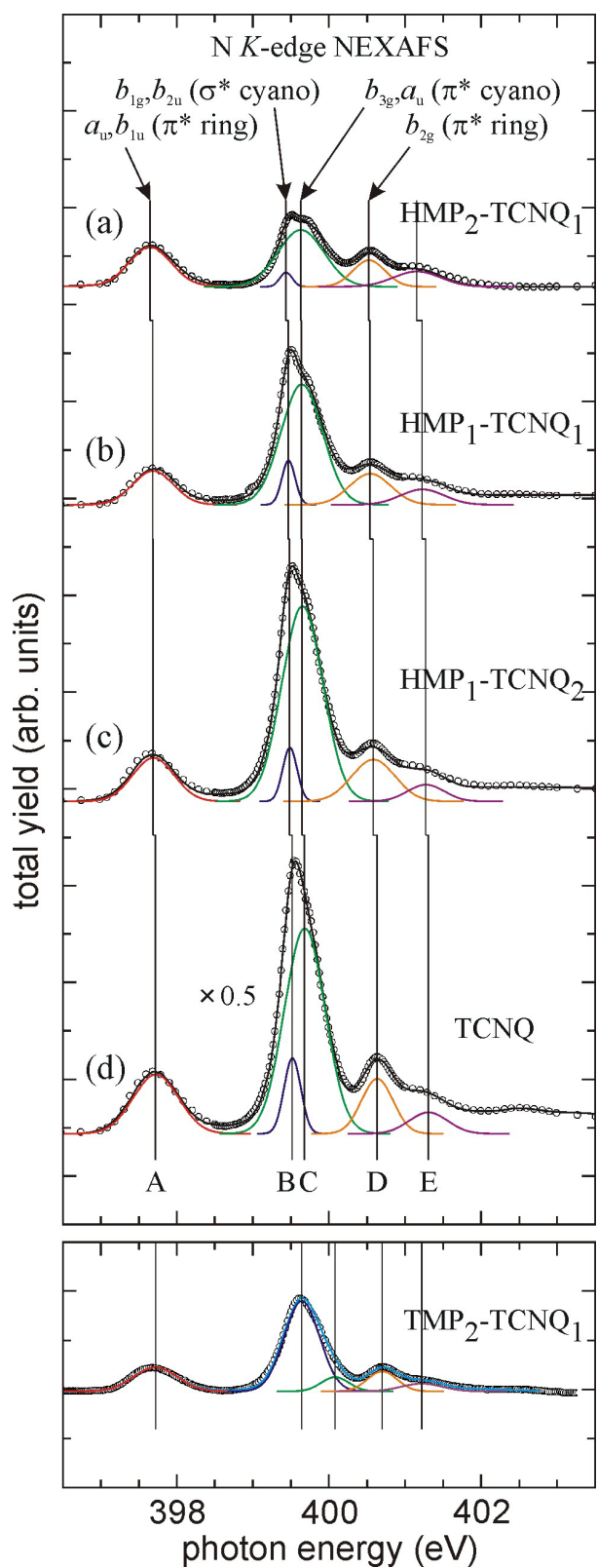

**Figure 2:** Nitrogen K-edge NEXAFS spectra of $HMP_x$-$TCNQ_y$ complexes with different stoichiometries. The yield scale is the same for all spectra; spectrum d was scaled by a factor of 0.5. Circles denote experimental data; thin curves below the spectra mark partial spectra of the transitions as obtained from a multi-peak fit routine; curves through the dots are the result of the fit (sum of the partial spectra). Bottom panel: Same for $TMP_2$-$TCNQ_1$.

## 3.2 Oxygen K-edge NEXAFS

Oxygen is only contained in the methoxy-groups of TMP and HMP. So its edge fine structure is a fingerprint of the donor moieties. Fig. 3 shows a series of spectra with f, g and h corresponding to the same samples as the series a, b and c in Fig. 2. Spectrum e has been taken for pure HMP. A number of signals (F-K) can be identified that partly show strong intensity changes for different stoichiometries.

The oxygen K-edge NEXAFS of the methoxy species chemisorbed on transition metal surfaces has been analyzed by Amemiya et al., we recall the peak assignments from [15]. The prominent pre-edge peak in the spectra (F) is separated by almost 3 eV from the next peak. This peak is located at 532.2 eV, in good agreement with the position of the lowest-lying peak measured for the methoxy species on Cu (531.7 eV [15]). This low-lying signal derives from the highest occupied molecular orbital (HOMO) 2e of the methoxy anion, being twofold degenerate and largely oxygen 2p-like (lone pair) with some hybridization with the antibonding π*-orbital of the C-O group. We denote transition F as ***O1s→2e[π*(C-O)]***. For multilayer methanol this signal is missing because the 2e orbital is filled [15]. For pure HMP (spectrum e) this signal is very weak, i.e. the 2e-orbital is completely filled. In other words, the methoxy group of the pure donor material HMP is fully anionic. However, when the orbital is not fully occupied the transition channel opens. This is obviously the case for the $HMP_2$-$TCNQ_1$ complex (spectrum f) and - to a weaker extent - also for the other two complexes (spectra g and h). The oscillator strength of transition F (essentially O1s→2p) is very high because it is dipole-allowed and characterized by a large overlap of initial- and final-state wavefunctions in the matrix element. This transition was also observed for the surface methoxy species on Cu and Ni. The data of [15] reveal that the 2e-vacancy of this species is quantitatively different on Ni(111) and Cu(111). For the latter case an effective electron population of 3.6*e* instead of 4*e* has been derived [16].

The second signal G at about 536 eV is associated with the transition ***O1s→6a₁[σ*(C-O)]***. Its intensity behaviour is similar to signal F. The methoxy σ*-resonances are observed also for multilayer films of methanol [15]. However, there is no π*-resonance for methanol [8]. The remaining signals H, I and K occur also for pure HMP and correspond to transitions into σ* and π*-orbitals of the aromatic ring system. The bottom panel of Figs. 2 and 3 show the spectra for the $TMP_x$-$TCNQ_y$ compound with 2:1 stoichiometry. This oxygen spectrum is the one with the highest O 1s → 2e resonance intensity found in this study.

The transition F is thus a perfect indicator of the ionicity of the methoxy group in the donor molecule. In Fig. 3 this signal increases strongly in the sequence of spectra h, g, f, i.e. for increasing HMP content. On the other hand, for pure HMP (spectrum e) the signal is missing. In the following section we will attribute the strong rise of signal F with increasing donor content in the compound to an according increase of the vacancy in the HOMO of the methoxy anion (2e orbital). In addition, the fit revealed small but significant shifts of the resonance positions in Fig. 3 towards higher photon energies, opposite to the shifts in Fig. 2.

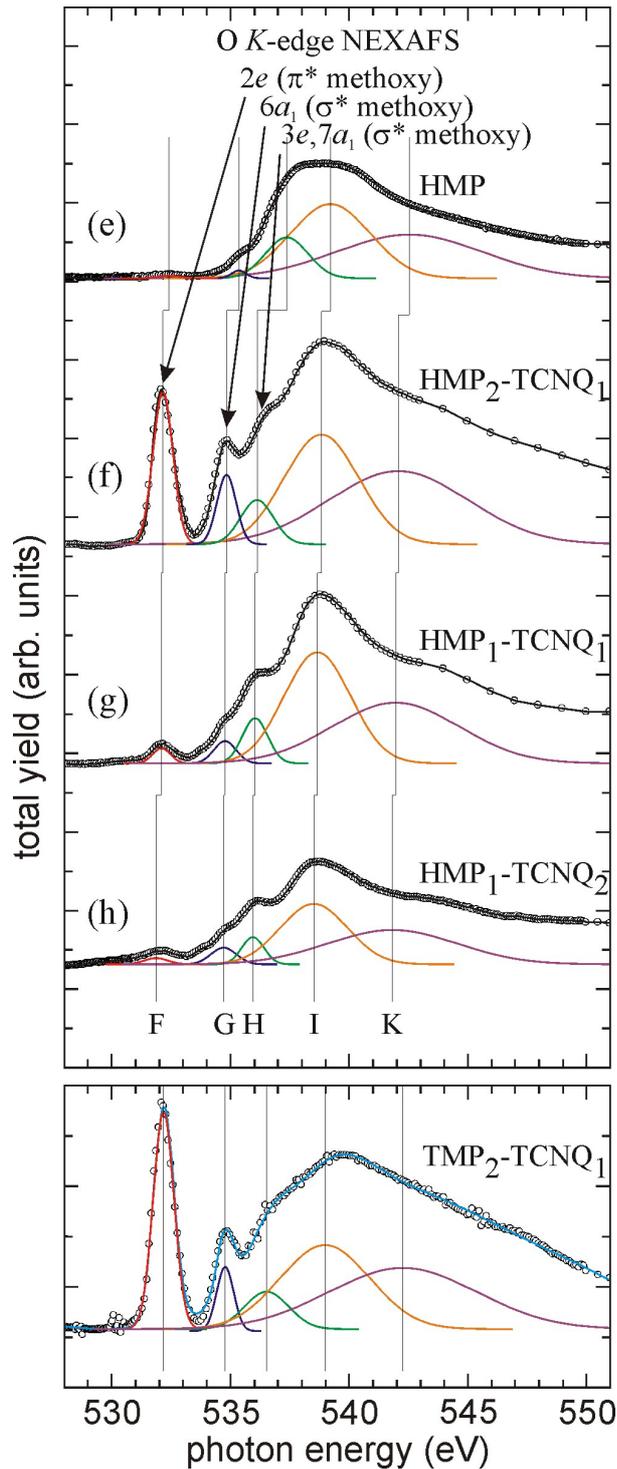

**Figure 3:** Same as Fig. 2 but for the K-edge NEXAFS spectra of oxygen. Bottom panel: Same for TMP$_2$-TCNQ$_1$.

### 3.3 Discussion

We now quantify the eye-catching result in the nitrogen and oxygen spectra (Figs. 2 and 3) that the intensities of several resonances vary strongly with increasing HMP content. Fig. 4 summarizes the intensity variations (a, b) and the energy shifts (c, d) of the oxygen and nitrogen resonances as function of the HMP/TCNQ ratio. The intensities have been determined as ratios of the areas under the corresponding fit curves in Figs. 2 or 3, normalized to the areas of peak A for nitrogen and peak I for oxygen (that are both not involved in the charge transfer). The increase of the intensities of resonances 2e (F) and $6a_1$ (G) in the oxygen spectra (Fig. 4a) for the 2:1 compound counteracts the intensity decrease of resonances $b_{3g}$ and $a_u$ (C) and $b_{1g}$ and $b_{2u}$ (B) in the nitrogen spectrum (Fig. 4b). These intensity variations with opposite behaviour for donor and acceptor give direct evidence of the participation of specific orbitals in the charge-transfer process in these compounds. The strong variations are the most direct indication of the intermolecular charge transfer. In particular the O 1s → 2e transition directly mirrors the charge depopulation in the HOMO of the methoxy moiety.

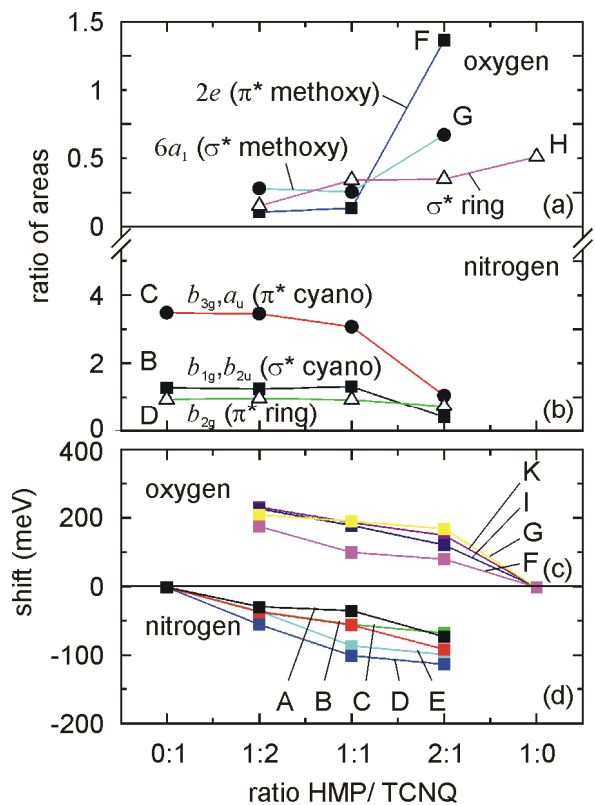

**Figure 4:** NEXAFS resonance intensities in the oxygen (a) and nitrogen spectrum (b) and energy shifts of the resonances in the oxygen (c) and nitrogen spectrum (d). Areas of the fitted peaks B, C and D are normalized to area A for nitrogen and peaks F, G and H are normalized to area I for oxygen. The energy shifts are referenced to the resonance positions in pure HMP and pure TCNQ.

A more indirect fingerprint is found in systematic shifts of the resonance positions as function of stoichiometry. Fig. 4c and d shows the energy shifts of the resonance positions for oxygen and nitrogen as determined from the fitted spectra. As reference we used the resonance positions of the pure donor and acceptor (stoichiometry 1:0 and 0:1, respectively). The shifts vary continuously with the HMP/TCNQ ratio and show an opposite sign for the donor and acceptor moiety. Since there are several transitions hidden in peak H, we did not include it in Fig.4c. The shifts are about a factor of 2 larger for the oxygen resonances than for the nitrogen signals.

Let us recall the different mechanisms contributing to shifts in the NEXAFS features (see, e.g. [8, 18]). The so-called valence shift reflects the redox state of the emitter atom: A change of the redox state changes the screening effect of the valence electrons on the nucleus. In turn, the binding energy of the electrons in the core level changes and the edge position appears shifted. In addition, a chemical shift is induced by different ligands, analogous to the chemical shift observed in XPS. Ligands with higher electronegativity reduce the effective charge density in the region of the excited atom and thus cause a shift of the NEXAFS resonances towards higher energies. Quantitatively, chemicals shifts are usually smaller in NEXAFS than in XPS, because the XPS probes the N→N-1 transition of the system, whereas in NEXAFS a transition to a neutral excited complex N→N* is observed, where both the core level and the unoccupied final state undergo a chemical shift. The shifts of initial and final states may partly cancel in the excitation energy.

The shift of all oxygen-related resonances towards higher photon energies (Fig. 4c) indicates an increasing deficiency of valence charge leading to reduced screening of the ion core. Vice versa, the shift of all nitrogen resonances to lower photon energies (Fig.4d) is indicative of an increasing charge density in the valence region. It is interesting to note that the variations in spectral weight of the resonances (Fig. 4a, b) and in the resonance positions (Fig. 4c, d) do not show the same quantitative behaviour: The spectral weight shows a sudden change between the 1:1 and 2:1 stoichiometry, whereas the shifts vary smoothly. This reflects the fact that the occupation of the final states of the NEXAFS transitions are affected by the partial charge transfer into specific orbitals, whereas the energetic positions vary because of the change in net charge of the whole molecule.

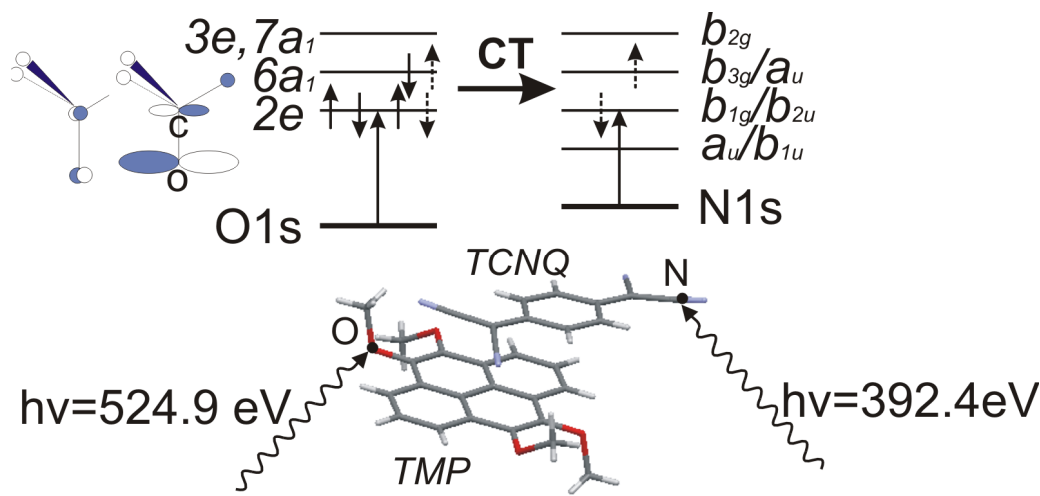

**Figure 5:** Transition scheme describing the action of the charge transfer from the donor to the acceptor. The NEXAFS spectra reveal 2e and $6a_1$ of the methoxy-group in TMP/HMP as donating orbitals and $b_{1g}$ and $b_{2u}$ of the cyano-group of TCNQ as the accepting orbitals.

A surprising fact is that the charge deficiency in the donor (HMP) is highest for the compound with maximum HMP content (2:1). The same is found for the $TMP_x$-$TCNQ_y$ compound with stoichiometry 2:1 (see lowest panel in Fig. 3). In a simple model the opposite effect might be expected. As the different stoichiometry leads to a different coordination and consequently to a different structure, we conclude that the relative arrangement of HMP and TCNQ may play an important role for the amount of charge transfer.

Fig. 5 summarizes the results in terms of a transition scheme including the charge transfer scenario (schematic). The partial charge transfer denoted by the fat arrow explains the intensity changes and energetic shifts of the observed NEXAFS resonances. Surprisingly, the states $6a_1$ as well as $b_{3g}$ and $a_u$ contribute to the charge transfer process although they are expected to be completely empty. This might be caused by the fact that NEXAFS does not probe the pure ground state properties. During the excitation all states are shifted to lower energies as a result of the core-hole attraction and therefore appear partially occupied.

## 5. Summary and Conclusion

Near-edge X-ray absorption fine structure (NEXAFS) spectroscopy was established as a tool to study the orbital specific charge transfer in novel donor-acceptor compounds. The novel functionalized pyrene-derivatives *4,5,9,10-tetramethoxypyrene* (TMP, $C_{20}H_{18}O_4$) and *2,4,5,7,9,10-hexamethoxypyrene* (HMP, $C_{22}H_{22}O_6$) were employed as donors in complexes with the classical strong acceptor *7,7,8,8-tetracyano-p-quinodimethane*, (TCNQ, $C_{12}N_4H_4$). $TMP_x$-$TCNQ_y$ and $HMP_x$-$TCNQ_y$ crystals in stoichiometries x:y = 1:2, 1:1 and 2:1 were grown by vapour diffusion of hexane into a dichloromethane solution of the components.

Oxygen and nitrogen K-edge NEXAFS spectra were exploited as local probes of the donor and acceptor molecules, respectively. Oxygen is only contained in the methoxy-moiety of the donor and nitrogen in the cyano-moiety of the acceptor. The orbital selectivity of the NEXAFS resonances allows to precisely elucidate the participation of specific orbitals in the charge-transfer process.

The spectra revealed partial charge transfer from the methoxy-orbitals 2e ($\pi^*$) and $6a_1$ ($\sigma^*$) of the donor to the cyano-orbitals $b_{3g}$ and $a_u$ ($\pi^*$) and - to a weaker extent - to $b_{1g}$ and $b_{2u}$ ($\sigma^*$) of the acceptor. In particular, the occupation of 2e (being the HOMO of the methoxy-species without charge transfer) reflects the anionic character of the methoxy-groups. Surprisingly, the charge transfer is maximum for the compounds with highest HMP or TMP contents. As additional indirect signature of a charge transfer, all spectral features of the donor and acceptor are shifted to higher and lower photon energies, respectively.

Providing quantitative access to the relative occupation of specific orbitals, the NEXAFS approach constitutes the most direct probe of the charge-transfer mechanism in organic salts found so far. Although demonstrated for the specific example of pyrene-derived donors with the classical acceptor TCNQ, the method is very versatile and can serve as routine probe for novel CT-complexes on the basis of functionalized polycyclic aromatic hydrocarbons.

**Acknowledgements**

The project is funded through Transregio SFB TR 49 (Frankfurt, Mainz, Kaiserslautern), Graduate School of Excellence MAINZ and Centre for Complex Materials (COMATT), Mainz. We thank C. Felser, S. Naghavi (Inst. for Inorganic and Analytical Chemistry, Univ. Mainz) and M. Huth, V. Solovyeva and M. Rudloff (Univ. of Frankfurt/Main) for fruitful cooperation. We acknowledge the ANKA Angströmquelle Karlsruhe for the provision of beamtime.